\documentclass[10pt,aps,pre,twocolumn,superscriptaddress]{revtex4-1}

\usepackage{graphicx}
\usepackage{amsmath, amssymb}
\usepackage{array,booktabs}

\usepackage{dcolumn,bm}

\usepackage[english]{babel}

\usepackage[mathscr]{euscript}

\usepackage{bm}
\usepackage{upgreek}

\usepackage{hyperref}
\usepackage{xcolor}

\graphicspath{{./figures/}}

\begin{document}

\title{Dynamics of sedimenting active Brownian particles}

\author{J\'er\'emy Vachier}\email{jeremy.vachier@ds.mpg.de}
\affiliation{Max Planck Institute for Dynamics and Self-Organization, Am Fa{\ss}berg 17, 37077 G\"ottingen, Germany}

\author{Marco G. Mazza}\email{marco.mazza@ds.mpg.de}
\affiliation{Max Planck Institute for Dynamics and Self-Organization, Am Fa{\ss}berg 17, 37077 G\"ottingen, Germany}
\affiliation{Interdisciplinary Centre for Mathematical Modelling and Department of Mathematical Sciences, Loughborough University, Loughborough, Leicestershire LE11 3TU, UK}

\begin{abstract}
We investigate the stochastic dynamics of one sedimenting active Brownian particle in three dimensions under the influence of gravity and passive fluctuations in the translational and rotational motion.
We present an analytical solution of the Fokker--Planck equation for the stochastic process which allows us to describe the dynamics of the active Brownian particle in three dimensions.   We address the time evolution of the monopole,  the polarization, and the steady-state solution. We also perform Brownian dynamics simulations 
and study the effect of the activity of the particles on their collective motion. 
These results qualitatively agree with our model. Finally, we compare our results with experiments [J. Palacci \emph{et al.}, Phys. Rev. Lett. \textbf{105}, 088304 (2010)] and find very good agreement.

\end{abstract}

\maketitle

\section{Introduction}

Active particles convert energy from chemical, biological, or other processes into motion. The study of active particles, and especially their collective motion, has received much attention due to a renewed interest in the physical principles underlying the motion of, \emph{e.g.}, plankton or bacteria, and also on account
 of technological applications involving both biological and artificial controllable active systems~\cite{elgetiRPP2015,ghoshNanoLett2009,kimAdvMat2013}. Active particles exhibit a fascinating multitude of interesting behaviors from the single particle to collective states~\cite{maassARCMP2016,kruegerPRL2016,jinPNAS2017}, 
due to their nonequilibrium nature. 

Typically, active particles move in an aqueous environment, where, because of their size,  viscous forces dominate, and inertial forces are completely negligible. In fact, consideration of the Navier--Stokes equations identifies that the nature of the dynamics is dictated by the ratio of viscous to inertial forces, known as the Reynolds number $\mathcal{R}=\sigma v\rho/\eta$, where $\sigma$ is the typical size of the microorganism, 
$v$ its mean velocity, and $\rho$, $\eta$ are the fluid's density and viscosity, respectively. For motile bacteria $\mathcal{R}\approx 10^{-5}$. 
As noted by Purcell~\cite{purcellAmJPhys1977}, this means that if the propulsion of the active particle were to suddenly disappear, it would only coast for $0.1$ \AA. Thus, the state of motion  is only determined  by the forces acting at that very moment, and inertia is negligible.

Even in dilute suspensions, where particle-particle interactions can largely be neglected, and the dynamics are dominated by the balance of active motion and gravity, interesting results are found~\cite{GolestanianPRL2009,Enculescu-PRL-2011,tenhagenNatComm2014,campbellJCP2017}. 
Palacci \emph{et al.}~\cite{Palacci-PRL-2010} showed experimentally with active Janus colloids that activity increases the sedimentation length, by increasing the effective diffusivity.
More recently, Ginot {\it et al.}~\cite{Ginot-PRX-2015} characterized  the equation of state of sedimenting  active colloids as a function of the activity. 

Theoretical studies of active particles, based on the framework of active Brownian particles~\cite{Romanczuk-EPJST-2012} and stochastic processes, have mostly focused on two-dimensional systems~\cite{Sevilla-PRE-2015,Pototsky-EPL-2012,Sevilla-PRE-2014,Wagner-JFM-2017,hermann-softmatter-2018}. A complete description in three dimensions (3D) in terms of the Fokker--Planck equation is challenging \cite{Wolff-EPJE-2013, Sevilla-PRE-2016} and some recent progress in the theory of one active particle \cite{Tailleur-EPL-2009, Enculescu-PRL-2011,Kurzthaler-SP-2016} highlights the fact that many questions are still open, especially in 3D.  For example, in dilute suspensions, what is the transient sedimenting dynamics? the emergence of polarization (and possibly higher orders) is intriguing and currently under investigation~\cite{ginot-Arxiv-2018}; what are the appropriate variables to construct an equation of state? In denser suspension, the important role of hydrodynamic interactions makes the situation even more complicated. For what physical conditions is the sedimenting steady state stable? What are the other possible steady states? Can we write an equation of state in this case? What are its relevant dynamical variables? 
In this work, we address the first question, that is, the transient state.\\
\indent We aim to analytically characterize the sedimentation of one active Brownian particle in 3D and, by means of Brownian dynamics simulations for many weakly-interacting particles. 
First, we analytically describe the sedimentation of one active particle under gravity with two overdamped Langevin equations  and the associated Fokker--Planck equation to obtain the particles' density profile in the direction of gravity. The density profile is obtained from the probability density function $P(\bm{r},\bm{e},t|\bm{r}_0,\bm{e}_{0},t_{0})$ of finding an active particle at the position $\bm{r}$, with an orientation $\bm{e}$ at time $t$, given the initial state $(\bm{r}_0,\bm{e}_{0},t_{0})$. Due to the complexity of the problem, finding the general expression of $P(\bm{r},\bm{e},t|\bm{r}_0,\bm{e}_{0},t_{0})$ in 3D is challenging. This method allows us to maintain coupling between the orientation and the position obtained in 3D, which we then specialize in one direction.  Furthermore,  in comparison with previous work~\cite{Tailleur-EPL-2009,Nash-PRL-2010,Enculescu-PRL-2011,Wolff-EPJE-2013,hermann-softmatter-2018} this method has the additional advantage of providing access to the full temporal dynamics, and is not limited to steady-state conditions, so that we can also investigate high P\'eclet numbers. We find an approximate solution for the time-dependent monopole, polarization and the steady-state solution.
Secondly, we perform Brownian dynamics simulations to describe the sedimentation of many particles, where fluid-mediated hydrodynamic interactions  are approximated via a short-range potential with up-down symmetry. \newline
The remainder of this work is organized as follows.
In section~\ref{sec:analytic}, we introduce  the stochastic process and solve the associated Fokker--Planck equation for a single, sedimenting active Brownian particle. In section~\ref{sec:simul}, we show the results of Brownian dynamics simulations of dilute suspensions of active particles. Finally, in section~\ref{sec:concl} we discuss our conclusions.

\section{Analytical solution for a single active Brownian particle}\label{sec:analytic}

\begin{figure}
\centering
\includegraphics[width=\columnwidth]{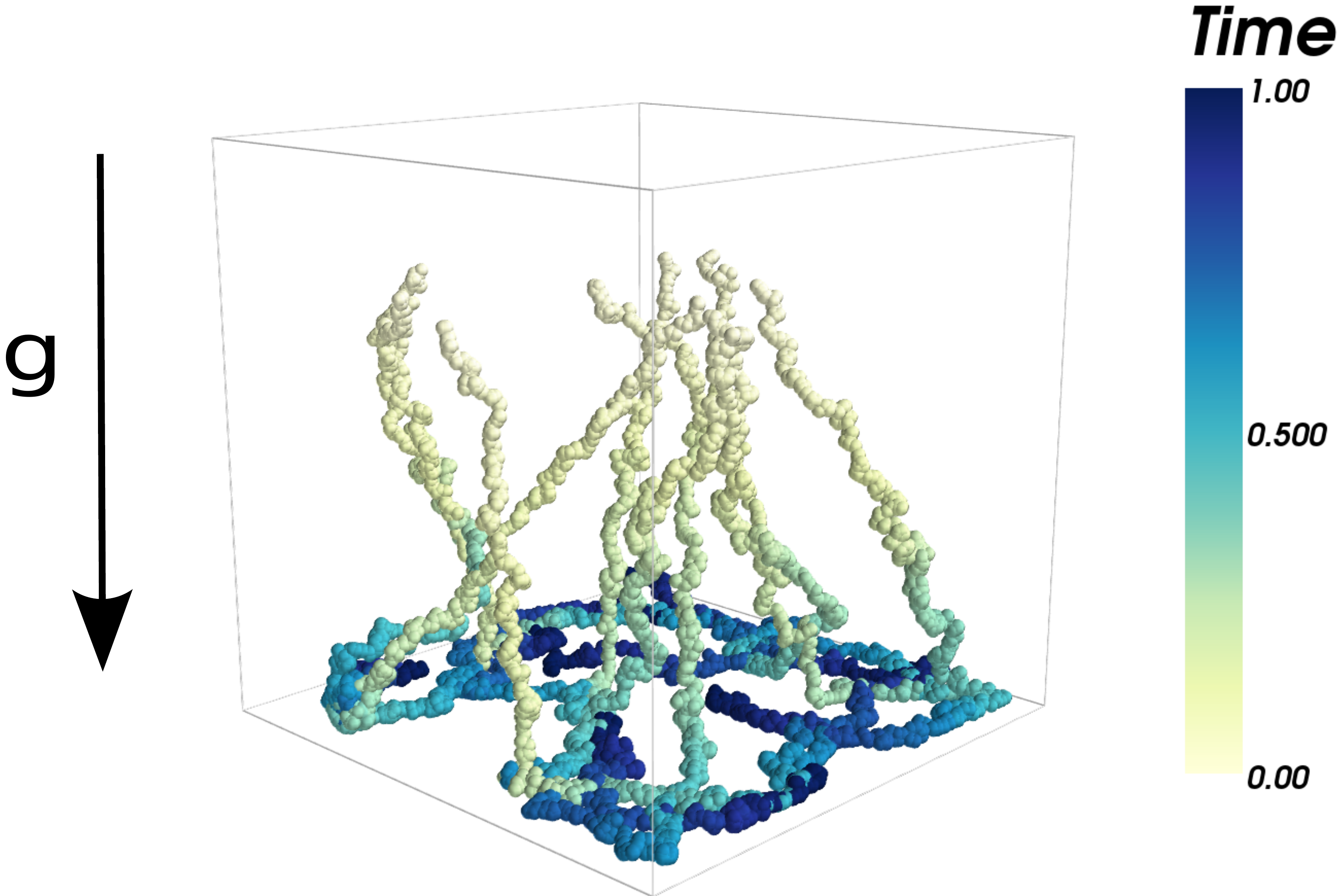}
\caption{Perspective view of the 3D motion of a few active Brownian particles under gravity in the presence of a reflective wall at the bottom. The trajectories are marked by showing the overlaid particles at subsequent times. The color code of each sphere indicates the time of that configuration.}
\label{globalpicture}
\end{figure}

We study analytically the motion of one self-propelled microscopic particle (active particle), considered as a point particle, in 3D under an external force: gravity. An example of this motion is shown in the fig.~\ref{globalpicture}. The activity of the particle means that it is able to convert energy in order to move. We represent the self-propulsion with a constant speed $v_{s}$ acting on the particle.
Typically, an active particle moves inside a fluid and due to its microscopic size we cannot neglect the influence of thermal fluctuations caused by the surrounding fluid buffeting the particle. The interactions with the fluid are represented by stochastic terms as for a Brownian particle. Due to gravity, the suspended active particle approaches a stationary state where its position has an increased probability of being close to the confining interface. This phenomenon is called sedimentation. To describe this motion, we derive a Fokker--Planck equation \cite{Van-Kampen-Stochastic-Processes, Gardiner, Risken,Frank-Nonlinear,Pavliotis-Grigorios-Stochastic-Processes}.

We treat the motion of one active particle, described as a point particle, in 3D under gravity by considering a single active Brownian particle moving with a constant active speed $v_s$ along a direction represented by the orientation $\bm{e}$, subject to random fluctuations. The motion of the active particle is biased by a drift velocity $- v_{g}\bm{z}$ in the direction of gravity. Our system is then described by two overdamped Langevin equations
\begin{align}
\frac{d}{dt}\bm{r}(t) & = v_{s}\bm{e}(t) - v_{g}\bm{z} + \bm{\xi}(t)\,, \label{Langevin equation1}\\ 
\frac{d}{dt}\bm{e}(t) &= \bm{\xi}_{\mathrm{e}}(t)\times\bm{e}(t)\,.
\label{Langevin equation2}
\end{align}
The random fluctuations are modeled in terms of the vectors $\bm{\xi}$ and $\bm{\xi}_{\mathrm{e}}$, with zero-mean, Gaussian white noise components, and with variance $\langle\xi_{i}(t)\xi_{j}(t')\rangle = 2 D_{\mathrm{t}}\delta_{ij}\delta(t-t')$,  $\langle\xi_{\mathrm{e}i}(t)\xi_{\mathrm{e}j}(t')\rangle = 2 D_{\mathrm{e}}\delta_{ij}\delta(t-t')$, where $D_{\mathrm{t}}$ and $D_{\mathrm{e}}$ are the translational and rotational diffusivities, respectively, $\delta_{ij}$ is the Kronecker delta, and $\delta(t)$ the Dirac distribution.

From eq.~\eqref{Langevin equation1}-\eqref{Langevin equation2} we can derive a Fokker--Planck equation that accounts for the evolution in time of the one-particle probability density function, $P(\bm{r},\bm{e},t|\bm{z}a,\bm{e}_{0},t_{0})$, of finding an active particle under gravity diffusing in 3D, with the initial condition $P(\bm{r},\bm{e},t = t_{0}|\bm{z}a,\bm{e}_{0},t_{0})=\langle\delta(\bm{r}-\bm{z}a)\delta(\bm{e}-\bm{e}_{0})\rangle$. In the following, to lighten the notation we will use $P(\bm{r},\bm{e},t)=P(\bm{r},\bm{e},t|\bm{z}a,\bm{e}_{0},t_{0})$.

 After some manipulation (see Appendix \ref{sec:app1}), we obtain the following Fokker--Planck equation
\begin{align}\label{FPE}	 
\frac{\partial}{\partial t} P(\bm{r},\bm{e},t) &= -v_{s}\bm{e}\cdot \nabla P(\bm{r},\bm{e},t) 
+ v_{g}\frac{\partial}{\partial z} P(\bm{r},\bm{e},t) \nonumber\\
&+ D_{\mathrm{t}}\nabla^{2}P(\bm{r},\bm{e},t) + D_{\mathrm{e}}\mathsf{L}_{\mathrm{e}}P(\bm{r},\bm{e},t)\,,	
\end{align}
 with  $\mathsf{L}_{\mathrm{e}}\equiv \left[ \frac{1}{\sin\theta} \frac{\partial}{\partial\theta}\left(\sin\theta  \frac{\partial}{\partial\theta} \right) + \frac{1}{\sin^2\theta} \frac{\partial^2}{\partial \phi^2}\right]$  the Laplace--Beltrami operator on the 2-sphere $\mathbb{S}^2$, and where we expressed $\bm{e}=(\sin\theta\cos\phi,$ $\sin\theta\sin\phi, \cos\theta)^\mathsf{T}$ in spherical coordinates. 
Equation~\ref{FPE} can be written symbolically as a continuity equation
\begin{equation}\label{eq:continuity}
\frac{\partial}{\partial t}P(\bm{r},\bm{e},t)=-\nabla \cdot \bm{J}\,,
\end{equation} 
 which defines the current $\bm{J}$.
To solve eq.~\eqref{FPE}, we proceed in the following way: (i) we will move to Fourier space; (ii) we will use an expansion in terms of eigenfunctions of the Fokker--Planck operator; (iii) we will perform a multipole expansion; (iv) we will focus on the dependence of the probability on the $z$-direction, along which gravity applies; and (v) we will perform the inverse-Fourier transform. 

The Fourier transform of eq.~\eqref{FPE} then reads
\begin{align}\label{eq:FPEFourier}
\frac{\partial}{\partial t}{\widehat{P}}(\bm{k},\bm{e},t) &= iv_{s}\bm{e}\cdot \bm{k} {\widehat{P}}(\bm{k},\bm{e},t) 
				-iv_{g}k_{z} {\widehat{P}}(\bm{k},\bm{e},t)\nonumber\\ 
			   &- D_{\mathrm{t}}k^{2}{\widehat{P}}(\bm{k},\bm{e},t) 
			   +D_{\mathrm{e}}\mathsf{L}_{\mathrm{e}}{\widehat{P}}(\bm{k},\bm{e},t)\,.
\end{align}

Let us for the moment consider the simple case  $v_{s}=0$ and $v_g=0$ which corresponds to a simple Brownian particle. Because the operator $\mathsf{O}_\mathrm{FP}=(\frac{\partial}{\partial t}+D_{\mathrm{t}}k^2-D_{\mathrm{e}}\mathsf{L}_{\mathrm{e}})$ is Hermitian, its eigenfunctions $e^{- D_{\mathrm{t}}k^{2}t}e^{-\lambda_{n}D_{\mathrm{e}} t}Y_{n}^{m}(\theta,\phi)$
form an orthonormal basis of the space of our solutions, where $\lambda_n=n(n+1)$,  $Y_{n}^{m}(\theta,\phi)=(-1)^m \sqrt{\frac{2n+1}{4\pi}\frac{(n-m)!}{(n+m)!}} P_{n}^{m}(\cos(\theta))e^{im\phi}$ are the spherical harmonics including the Condon--Shortley phase factor, and $n, m\in \mathbb{N}$,  $-n\leq m\leq n$. Additionally, because $\mathsf{O}_g=iv_{g}k_{z}$ is simply a multiplicative scalar, an eigenfunction of $\mathsf{O}_\mathrm{FP}+\mathsf{O}_g$ is $e^{-(iv_gk_z+ D_{\mathrm{t}}k^{2})t}e^{-\lambda_{n}D_e t}Y_{n}^{m}(\theta,\phi)$.

Taking into account the initial condition 
\begin{align*}
{\widehat{P}}(\bm{k},\bm{e},t = t_{0}) &=\frac{1}{(2\pi)^{\frac{3}{2}}}\int d^3\bm{r}e^{-i\bm{k}\cdot\bm{r}}{P}(\bm{r},\bm{e},t= t_{0})\\
&=\frac{1}{(2\pi)^{\frac{3}{2}}}e^{-ik_{z}a}
\end{align*}
 and the linearity of eq.~\eqref{eq:FPEFourier}, we will search for solutions of the form

\begin{align}\label{eq:gen-expansion}
{\widehat{P}}(\bm{k},\bm{e},t)&=  e^{-\left(iv_{g}k_{z} + D_{\mathrm{t}}k^{2}\right)t}e^{-ik_{z}a}\nonumber\\
&\sum_{n=0}^{+\infty}\sum_{m=-n}^{+n}\widehat{P}_{n}^{m}(\bm{k},t)
e^{-D_{\mathrm{e}}n(n+1)t}Y_{n}^{m}(\bm{e})\,,
\end{align}
where the coefficients $\widehat{P}_{n}^{m}(\bm{k},t)$ are determined by imposing that the expression in eq.~\eqref{eq:gen-expansion} satisfy eq.~\eqref{eq:FPEFourier} (see Appendix \ref{sec:app2}). Physically, the infinite sums on the right-hand side of eq.~\eqref{eq:gen-expansion} represent the increasingly faster decay with time of higher-order spherical harmonics~\cite{Sevilla-PRE-2016}.

Because of the rotational dynamics in our problem, it is convenient to explicitly highlight the underlying physical symmetries by expanding the full probability $P(\bm{r},\bm{e},t)$ in terms of spherical tensors, that is, the irreducible representations of the rotation operator. Each spherical tensor transforms like the eigenfunctions of the angular momentum of corresponding rank $n=0,1,2,\dots$, where the first three tensors represent the density $\rho$ (monopole, $n=0$), the polarization $\bm{D}$ (dipole, $n=1$), and the nematic tensor $\mathbf{Q}$ (quadrupole, $n=2$), respectively. The probability can then be expanded as
\begin{align}\label{eq:multipole}
P(\bm{r},\bm{e},t) = \rho(\bm{r},t) + \bm{D}(\bm{r},t)\cdot \bm{e} + \bm{e}\cdot \mathbf{Q}(\bm{r},t)\cdot \bm{e}+ \dots
\end{align}
In the large time limit, the monopole term $\rho(\bm{r},t)$ will dominate the sedimentation process (while higher order terms in eq.~\eqref{eq:multipole} are relevant for observables with shorter characteristic time scales). Its Fourier transform can be found by truncating the sum in eq.~\eqref{eq:gen-expansion} at the $n=0$ term, that is 
\begin{equation}
\hat{\rho}(k_{z},t)=\frac{1}{\sqrt{4\pi}} e^{-\left(iv_{g}k_{z}+D_{\mathrm{t}}k_{z}^{2}\right)t}e^{-ik_{z}a}\widehat{P}^{0}_{0}\,.
\end{equation}
After some computations (see Appendix \ref{sec:app3}), we can work out  the equation governing the dynamics of ${\widehat{P}}^{0}_{0}$
\begin{equation}
\frac{\partial^2}{\partial t^2} {\widehat{P}}^{0}_{0} + 2D_{\mathrm{e}}\frac{\partial}{\partial t}{\widehat{P}}^{0}_{0} + \frac{v_{s}^{2}}{3}k_{z}^{2}{\widehat{P}}^{0}_{0} = 0\,,
\label{eq:Telegrapher}
\end{equation}
which is the telegrapher's equation~\cite{ilyinCMP2013}, and accounts for processes with a finite speed of propagation.

A solution of eq.~\eqref{eq:Telegrapher} reads
$\widehat{P}^{0}_{0}(k_{z},t)=e^{-D_{\mathrm{e}} t}[\widehat{F}(k_{z})e^{-iw(k_{z})t}+\widehat{G}(k_{z})e^{iw(k_{z})t}]$,
where $w(k_{z})=(k_{z}^{2}\frac{v_{s}^{2}}{3}-D_{\mathrm{e}}^{2})^{1/2}$, with $\widehat{F}(k_{z})$ and $\widehat{G}(k_{z})$ arbitrary functions of the wavevector in the $z$-direction $k_z$. The expression for the monopole is found from the inverse Fourier transform, and reads
\begin{align*}
\rho(z,t) &= \int\limits_{-\infty}^{+\infty}\frac{dk_{z}}{\sqrt{2\pi}}\hat{\rho}(k_{z},t)e^{ik_{z}z}\nonumber\\
	&=\frac{e^{-D_{\mathrm{e}} t}}{\pi\sqrt{8}}\int\limits_{-\infty}^{+\infty}dk_{z}e^{-iv_{g}k_{z}t-D_{\mathrm{t}}k_{z}^{2}t}e^{-ik_{z}a}e^{ik_{z}z}\nonumber\\
	&\times[\widehat{F}(k_{z})e^{-iw(k_{z})t}+\widehat{G}(k_{z})e^{iw(k_{z})t}]\,.
\end{align*}
The term $w(k_{z})=({k_{z}^{2}{v_{s}^{2}}/{3}-D_{\mathrm{e}}^{2}})^{1/2}$  in the exponential makes it difficult to perform the inverse Fourier transform. However, because we are interested in the long-wavelength limit of the sedimentation profile, it is natural to consider a Taylor expansion of $w(k_{z})$ around $k_{{z}}=0$
\begin{align*}
\rho(z,t)&= \frac{e^{-D_{\mathrm{e}} t}}{\pi\sqrt{8}}\int\limits_{-\infty}^{+\infty}dk_{z}e^{-D_{\mathrm{t}}k_{z}^{2}t-iv_{g}k_{z}t}e^{ik_{z}z}e^{-ik_{z}a}\nonumber\\
	& \times \left[\widetilde{F}e^{-D_{\mathrm{e}}t+\frac{v_{s}^{2}k_{z}^{2}t}{6D_{\mathrm{e}}}}+\widetilde{G}e^{D_{\mathrm{e}}t-\frac{v_{s}^{2}k_{z}^{2}t}{6D_{\mathrm{e}}}}\right]\,.
\end{align*}
where $\widetilde{F}$ and $\widetilde{G}$ are defined as follow
\begin{align*}
  \widehat{F}(k_{z})e^{-iw(k_z)t}
& = \widehat{F}(k_{z})e^{-D_{\mathrm{e}}t+\frac{v_{s}^{2}k_{z}^{2}t}{6D_{\mathrm{e}}}-iO((w(k_z)t)^2)} \\
& = e^{-D_{\mathrm{e}}t+\frac{v_{s}^{2}k_{z}^{2}t}{6D_{\mathrm{e}}}}\widehat{F}(k_{z})e^{-iO((w(k_z)t)^2)}\\
& = e^{-D_{\mathrm{e}}t+\frac{v_{s}^{2}k_{z}^{2}t}{6D_{\mathrm{e}}}}\widetilde{F}
\end{align*}
and similarly for $\widetilde{G}$.
Elementary integration yields
\begin{align}\label{mainpdf}
\rho(z,t) &= \frac{1}{\sqrt{8\pi}} \Bigg[e^{-2D_{\mathrm{e}}t} \frac{\widetilde{F}}{\sqrt{D_\mathrm{eff}^- t}}e^{-{(z-a-v_{g}t)^{2}}/({4 D_\mathrm{eff}^- t})}\nonumber\\
&+\frac{\widetilde{G}}{\sqrt{D_\mathrm{eff}^+ t}}e^{-{(z-a-v_{g}t)^{2}}/({4D_\mathrm{eff}^+ t})}\Bigg]\,,
\end{align}
where we have defined the effective diffusivities $D_\mathrm{eff}^\pm\equiv D_{\mathrm{t}}\pm \frac{v_{s}^{2}}{6D_{\mathrm{e}}}$. That active motion enhances diffusion has been repeatedly observed in experimental \cite{Palacci-PRL-2010} and theoretical works \cite{Kurzthaler-SP-2016,stark-EPJST-2016}. 
By imposing that mass is conserved during the sedimentation process
\[
\frac{d}{dt}\int dz \rho(z,t) = 0\,,
\]
we can determine the functions $\widetilde{F}=\exp(2D_et)$ and $\widetilde{G}=1$.

{\color{black}
In order  to describe sedimentation, we need to impose a reflective boundary condition  due to the confining wall located at $z=0$.  
By integrating \eqref{FPE} over the orientation, we find
\[
\frac{\partial}{\partial t}\rho(z,t)=-\frac{1}{2}v_s\frac{\partial}{\partial z}D(z,t)+v_g\frac{\partial}{\partial z}\rho(z,t)+D_t\frac{\partial^2}{\partial z^2}\rho(z,t)\,.
\]
The associated continuity equation reads
\[
\frac{\partial}{\partial t} \rho(z,t) = -\frac{\partial}{\partial z}J_z\,,
\]
and by imposing no flux $J_z=0$ at the wall, the boundary condition, of the Robin type, reads \cite{cates-epl-2013}
\begin{equation}
\left[D_t\frac{\partial}{\partial z}\rho(z,t)+v_g\rho(z,t)-\frac{1}{2}v_s D(z,t)\right]_{z=0}=0\,.
\end{equation}
Because of the large time limit, the monopole $\rho(z,t)$ dominate the sedimentation process and therefore the dipole $D(z,t)$ is negligible. Hence,
\begin{equation}
\left[D_t\frac{\partial}{\partial z}\rho(z,t)-v_g\rho(z,t)\right]_{z=0}=0\,.
\label{boundary}
\end{equation}
We rewrite eq.\eqref{mainpdf} as
\[
\rho(z,t)=\rho_1(z,t)+\rho_2(z,t)\,,
\]
where 
\[
\rho_1(z,t)= \frac{1}{\sqrt{8\pi}}\frac{e^{-(z-a-v_gt)^2/4D_\mathrm{eff}^-t}}{\sqrt{D_\mathrm{eff}^-t}}\,,
\]
and
\[
\rho_2(z,t)= \frac{1}{\sqrt{8\pi}}\frac{e^{-(z-a-v_gt)^2/4D_\mathrm{eff}^+t}}{\sqrt{D_\mathrm{eff}^+t}}\,.
\]
In order to impose no net flux across the reflective wall, we use the method of images, but as known in the theory of partial differential equation [Sommerfeld] the appropriate image system consists  of replacing the wall at $z=0$ with a mirror source placed at $z=-a$ (in addition to the real source at $z=a$) and a continuous sequence of images which take place at all points $\xi<-a$ \cite{redner2001,sommerfeld-1949-partial,aleksandrov-1999-mathematics,cox-2017-theory,chandrasekhar-1943-stochastic}. 
We can rewrite the probability density as
\begin{equation}
\rho_\mathrm{r}(z,t) = \rho(z,t|a)+A\rho(z,t|-a)+\int\limits^{-a}_{-\infty}k(\xi)\rho(z,t|\xi)d\xi\,,
\label{densitywallmanu}
\end{equation}
By applying eq.\eqref{densitywallmanu} to our system, $\rho_\mathrm{r}(z,t)$ reads
\begin{align}
\rho_r(z,t) &= \rho_{r1}(z,t|a)+\rho_{r2}(z,t|a) + A_1\rho_{r1}(z,t|-a)\nonumber\\
&+A_2\rho_{r2}(z,t|-a)+ \int\limits_{-\infty}^{-a}k_1(\xi)\rho_{r1}(z,t|\xi)d\xi  \nonumber\\
 &+ \int\limits_{-\infty}^{-a}k_2(\xi)\rho_{r2}(z,t|\xi)d\xi \,, 
\end{align}
where  the coefficients $A_1$, $A_2$, $k_1(\xi)$, and $k_2(\xi)$ are also found via the Robin boundary condition (see appendix \ref{sec:app4}) the solution yields
\begin{align}
&\rho_r(z,t)= \frac{v_g}{D_\mathrm{eff}^-\sqrt{2}} \mathrm{erfc}\left(\frac{z+a-v_gt}{2\sqrt{D_\mathrm{eff}^-t}}\right)e^{-v_gz/D_\mathrm{eff}^-}\nonumber\\
&+\frac{1}{\sqrt{8\pi}\sqrt{2D_\mathrm{eff}^-t}}\left[e^{\frac{-(z-a)^2}{4D_\mathrm{eff}^-t}}+e^{\frac{-(z+a)^2}{4D_\mathrm{eff}^-t}}\right]e^{\frac{-v_g(z-a)}{2D_\mathrm{eff}^-}-\frac{v_g^2t}{4D_\mathrm{eff}^-}}\nonumber\\
&+\frac{v_g}{D_\mathrm{eff}^+\sqrt{2}} \mathrm{erfc}\left(\frac{z+a-v_gt}{2\sqrt{D_\mathrm{eff}^+t}}\right)e^{-v_gz/D_\mathrm{eff}^+}\nonumber\\
&+\frac{1}{\sqrt{8\pi}\sqrt{2D_\mathrm{eff}^+t}}\left[e^{\frac{-(z-a)^2}{4D_\mathrm{eff}^+t}}+e^{\frac{-(z+a)^2}{4D_\mathrm{eff}^+t}}\right]e^{\frac{-v_g(z-a)}{2D_\mathrm{eff}^+}-\frac{v_g^2t}{4D_\mathrm{eff}^+}} \,.\label{reflectivedensity}
\end{align}
The steady state regime is given by taking the limit $t\to\infty$ of eq.\eqref{reflectivedensity} and reads
\begin{equation}
\lim\limits_{t\to\infty}\rho_r(z,t)=\frac{2v_g}{D_\mathrm{eff}^-\sqrt{2}}e^{-v_gz/D_\mathrm{eff}^-} +\frac{2v_g}{D_\mathrm{eff}^+\sqrt{2}}e^{-v_gz/D_\mathrm{eff}^+}
\label{steadystatef}
\end{equation}
}

In the following, we take the active particle's diameter $\sigma$, mass $m$ and its translational diffusion coefficient $D_{\mathrm{t}}$ as the units of length, mass and diffusivity. 
Thus, we can measure rotational diffusivity in terms of  $\widetilde{D}_{e}=D_{\mathrm{t}}/\sigma^{2}$.
A dimensionless measure of the relative strength of the self-propulsion to the diffusive behavior, that is, the relative persistence of the active motion, is given by the  P\'eclet number $\mathcal{P} = v_{s}\sigma/D_{\mathrm{t}}$.

\begin{figure}
\centering
\includegraphics[width=1.0\columnwidth]{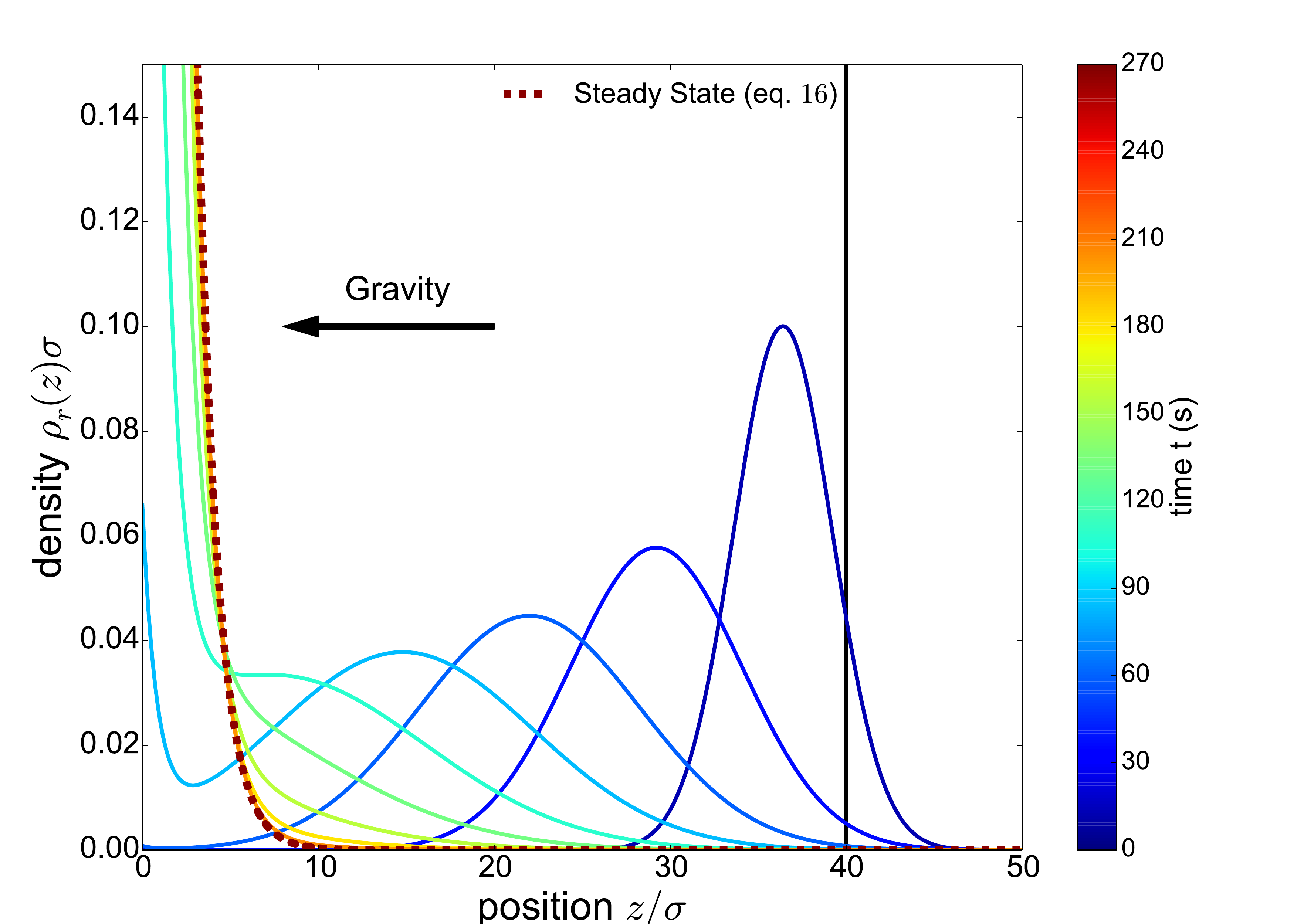}
\caption{{\color{black}Analytical sedimentation profile. Dependence of the density $\rho_{\mathrm{r}}(z,t)$ on the position $z$, computed from eq.~\eqref{reflectivedensity}, at different times for a system with one active Brownian particle under gravity in a cubic box of linear size $L=50\sigma$ with a reflective wall on bottom ($z=0$) and with gravity pointing in negative $z$-direction. At $t=0$, the initial position of the active particle is at $z/\sigma=40$, and the corresponding probability density is a Dirac delta distribution. With time  we observe a spreading of the density profile and a shift in the direction of gravity. In the steady state regime, eq. \eqref{steadystatef}, we obtain a sedimentation profile characterized by an exponential decay with distance. Different curves correspond to different instants during the time evolution. The model parameters are $v_{s}/v_{g} = 1.1$, and $D_{\mathrm{e}} {\sigma^{2}}/{D_{\mathrm{t}}} = 1.8$.}\label{withboundary}}
\end{figure}

\begin{figure}
\centering
\includegraphics[width=1.0\columnwidth]{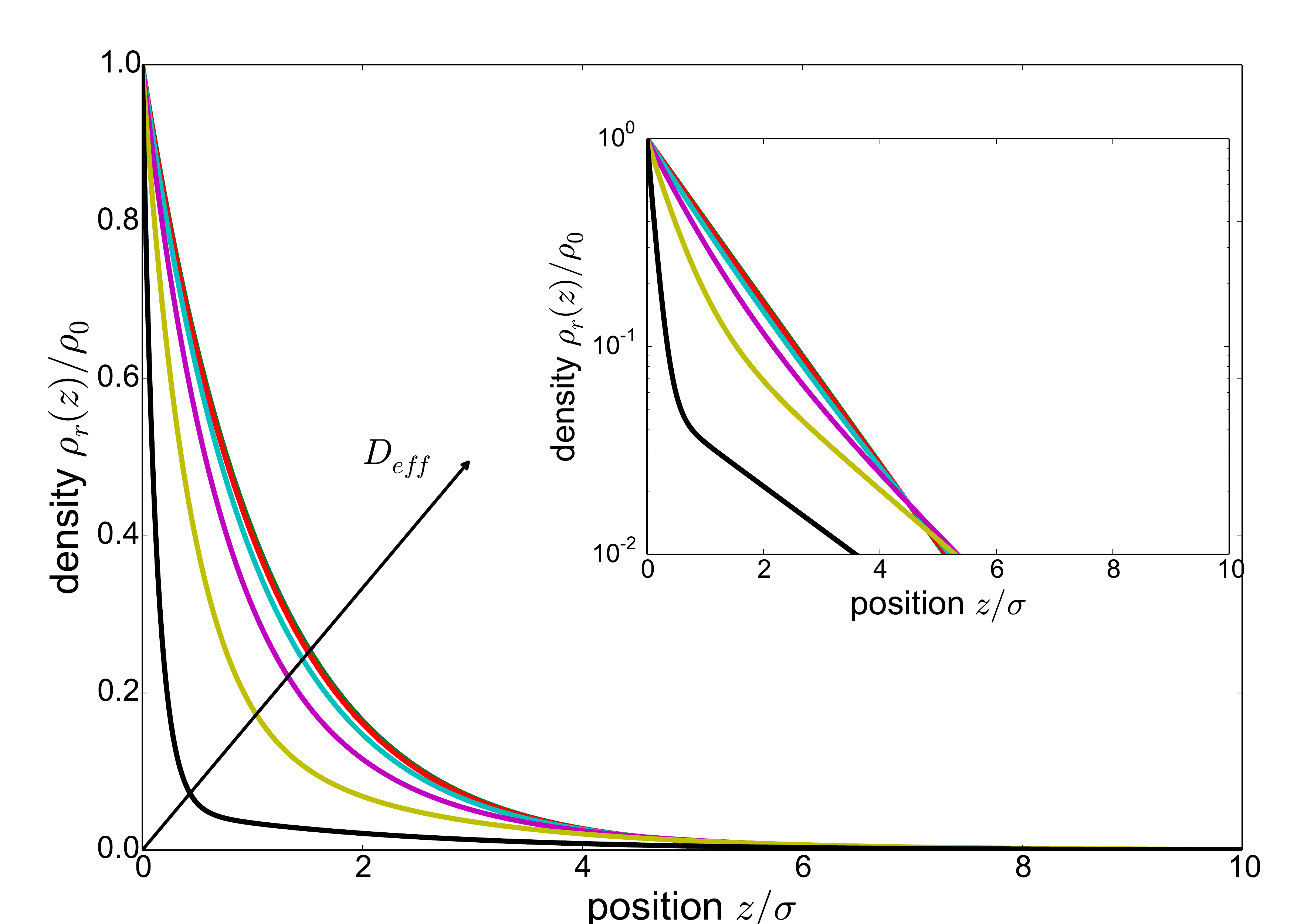}
\caption{{\color{black}Normalized sedimentation profile in the steady state regime for a reflective barrier. Dependence of the normalized density $\rho_{\mathrm{r}}(z)/\rho_{0}$ on the position $z$, in the steady state regime given by eq.~\eqref{steadystatef}.  Different curves correspond to the long time behavior of the sedimentation process for different values of the effective diffusion coefficient $D_\mathrm{eff}^\pm=D_{\mathrm{t}}\pm\frac{v_{s}^{2}}{6D_{\mathrm{e}}}$. We  observe a good match with the experimental  results in \cite{Palacci-PRL-2010}.  The  model parameters are $D_{\mathrm{e}} {\sigma^{2}}/{D_{\mathrm{t}}} = 1.8$, and $v_{s}/v_{g}\in [1.2,6.6]$.}\label{diffcoefstrength1}}
\end{figure}

Figure~\ref{withboundary} shows the evolution of the density profile $\rho_{\mathrm{r}}(z,t)$ found from our solution to the sedimentation process in eq.~\eqref{reflectivedensity}. 
The initial position  of the active particle is chosen at $z/\sigma=40$. The corresponding initial density $\rho_{\mathrm{r}}(z,t=0)$ is a Dirac delta distribution. As time progresses, we observe the shifting and flattening of the density profile. 
The steady state regime, given by eq.\eqref{steadystatef} is characterized by an exponential decay, which match the sedimentation profile.
We match our parameters with the experimental values given in \cite{Palacci-PRL-2010}, where the P\'eclet number $0.5 \lesssim\mathcal{P}\lesssim 5$, and we find a near-quantitative agreement with the experiments. 
As predicted in theoretical works~\cite{Wolff-EPJE-2013, Tailleur-EPL-2009}, $\rho_{\mathrm{r}}(z,t)$ decays exponentially away from the confining surface. Upon increasing the self-propulsion $v_s$, and therefore the effective diffusivity $D_{\mathrm{eff}}$, the density profile tends to spread away from the wall as observed in the experiment \cite{Palacci-PRL-2010}. This behavior is shown in fig.~\ref{diffcoefstrength1}.

The sedimentation length $\delta_\mathrm{eff}$ is the characteristic length scale of the decay of $\rho_{\mathrm{r}}(z,t)$ with $z$. It was found  to depend strongly on the activity of the self-propelling particle~\cite{Palacci-PRL-2010,Wolff-EPJE-2013}. 
 In general, we find a linear relationship governing the growth of $\delta_\mathrm{eff}$  with $D_\mathrm{eff}/v_g$, $\delta_\mathrm{eff}=c_0+D_\mathrm{eff}/v_g$. The constant $c_0\equiv c_0(v_g)$, and can be chosen to be zero, which is the value consistent with the experiments in~\cite{Palacci-PRL-2010}. The relationship between $\delta_\mathrm{eff}$ and $D_\mathrm{eff}$ provides a connection between the microscopic behavior of the active particle and the long-time emergent dynamics ~\cite{Palacci-PRL-2010}.
The precise nature of the density profile in proximity of the confining surface will be affected by a number of effects such as: electrostatics, and hydrodynamic interactions of the active particles with the walls. For example, in a recent work \cite{das-natcom-2015}, the authors show that boundaries can steer Janus colloids, which, as a result, move above the boundary at a fixed distance. These effects are not taken into account here.

Additional information about the active sedimentation process can be gained by considering the next term in the expansion eq.~\eqref{eq:multipole}, \emph{i.e.} the polarization. The probability density function becomes
\[
P(z,\cos(\theta),t) \simeq \rho(z,t) + D(z,t)\cos(\theta)\,.
\]
We can express the polarization $D$ by means of the Legendre polynomials. Again, we are only interested in the $z$-direction. In Fourier space we find
\begin{equation}
D(k_z,t)=\sqrt{\frac{3}{4\pi}}e^{-(iv_gk_z-D_tk_z^2)t}e^{-ik_za}\widehat{P}^{0}_{1}\,. 
\end{equation}
After some computations and applications of the boundary conditions (at $z=0$, $J = 0$ and $\theta=\pi$), the probability density function reads
\begin{align}
&P(z,\cos(\theta),t) = \frac{1}{\sqrt{8\pi}}\Bigg[\frac{1}{\sqrt{D_\mathrm{eff}^- t}}e^{-{(z-a-v_{g}t)^{2}}/({4 D_\mathrm{eff}^- t})}\nonumber\\
&+\frac{1}{\sqrt{D_\mathrm{eff}^+ t}}e^{-{(z-a-v_{g}t)^{2}}/({4D_\mathrm{eff}^+ t})}\Bigg]\nonumber\\
&+\sqrt{\frac{3}{4\pi}}\frac{\cos(\theta)}{\sqrt{2tD_t}}
\left(\frac{B_1+B_2}{C}+1\right)\cos(\alpha)
\nonumber\\
&\times e^{-(a^2-2az+z^2+2atv_g-2tzv_g+t^2v_g^2
-ft^2D_e^2v_s^2)/(4D_t t)}.
\label{pdfpolar}
\end{align}
\noindent We refer the readers to appendix \ref{sec:app5} for further details, and for the definition of the constants $B_1$, $B_2$, $C$, $\alpha$, and $f$.

\begin{figure}
\centering
\includegraphics[width=1.0\columnwidth]{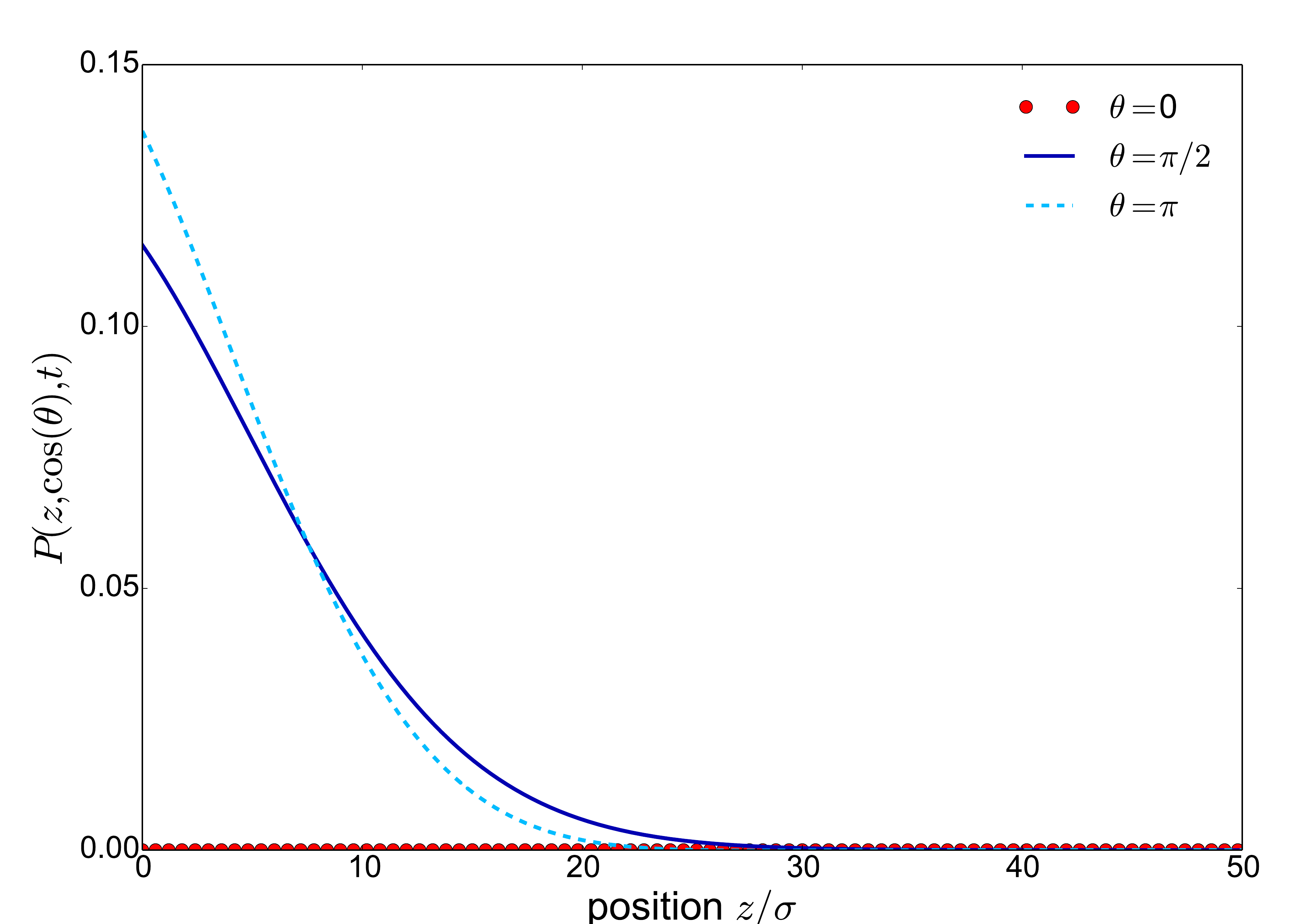}
\caption{Polarization. Dependence of the probability density function $P(z,\cos(\theta),t)$ on the position $z$, computed at large $t$ for three values of the orientation $\theta\in\{0,\pi/2,\pi\}$. The model parameters are $v_{s}/v_{g} = 1.1$, and $D_{\mathrm{e}} {\sigma^{2}}/{D_{\mathrm{t}}} = 1.8$.}\label{polarorder}
\end{figure}

After imposing the condition $P(z,\cos(\theta),t)\geq 0$, we show in fig.~\ref{polarorder} $P(z, \cos(\theta),t)$, at large $t$ and for three values of the orientation $\theta\in\{0,\pi/2,\pi\}$. 
As predicted in \cite{Enculescu-PRL-2011}, we observe an accumulation of active particles with a net polarization at the bottom wall, moving againts the wall ($\theta=\pi$). The qualitative picture is in agreement with \cite{Enculescu-PRL-2011}.

\section{Simulations of the collective motion}\label{sec:simul}


We next investigate a many-particle system composed of active Brownian particles under the effect of gravity. An analytical approach is a formidable task; thus we turn to numerical simulations. Even in the simpler case of passive Brownian particles, sedimentation is a complex process on account of velocity correlations and hydrodynamic interactions~\cite{batchelor-JFM-1972,Segre-PRL-1997,Piazza-RepProgPhys-2014}. Because we are interested in the impact of active motion on the sedimentation, we can reduce the nonlinearities associated to hydrodynamic interactions by working in the dilute limit, similarly to \cite{Palacci-PRL-2010,Enculescu-PRL-2011,Tailleur-EPL-2009,Wolff-EPJE-2013}. We do however consider weak hydrodynamic interactions via a short-ranged effective potential (see below).
 We perform Brownian dynamics simulations in 3D.

We consider again a typical Reynolds number
  $\mathcal{R} \ll 1$. An example of such particles in a biological
  setting is the microalga {\it Chlamydomonas reinhardtii}, which has
  a typical length $\sigma = 10~\upmu$m  and self-propulsion speed
  $v_{s}=60~\upmu$m s$^{-1}$. In colloidal physics  active Janus
  particles have a typical linear size $\sigma = 1~\upmu$m and
  self-propulsion speed which can vary as a function of the chemical gradient. As a reference, we can take the results found in experiment \cite{Palacci-PRL-2010}, where the self-propulsion  $v_{s}=(0.3-4)\upmu$m s$^{-1}$.

We describe the system by using two first-order stochastic differential equations, for $N$ active particles
\begin{align}
\frac{d}{dt}\bm{r}_{i}  &= v_{s}\bm{e}_{i} -\nabla\phi_{WCA} -v_{g}\bm{\widehat{z}}+ \bm{\xi}_{i}\,, \label{eq:simulation-1}\\ 
\frac{d}{dt}\bm{e}_{i} &= \bm{\xi}_{e_{i}}\times\bm{e}_{i} -\gamma\frac{\partial U}{\partial \bm{e}_{i}}\,, \label{eq:simulation-2}
\end{align}
where $i =1, \dots, N$, $\|\bm{e}_{i}\|=1$ (implemented by means of a Lagrangian multiplier), $v_{g}$ is the limiting velocity of a particle in the fluid under gravitational acceleration. 
In eq.~\eqref{eq:simulation-1}, $\phi_{WCA}=4\epsilon [({\sigma}/{r_{ij}})^{12}-({\sigma}/{r_{ij}})^{6}]+\epsilon$ is the Weeks--Chandler--Anderson potential~\cite{weeks-JCP-1971}, $r_{ij}=|\bm{r_{i}}-\bm{r_{j}}|$, representing a hard-core repulsion between active particles, where $\sigma$ is the linear size of the active particles, and $\epsilon$ is the energy scale of the repulsive interaction.
In eq.~\eqref{eq:simulation-2}, 
$U=\sum_{i\neq j}\cos^{2}(\theta_{ij})$ is the Lebwohl--Lasher potential, which we use to model to first approximation the up-down symmetric  interaction due to hydrodynamics that tends to align neighboring particles (see \emph{e.g.} \cite{BaskaranPNAS2009}).

\begin{figure}
\centering
\includegraphics[width=1.0\columnwidth]{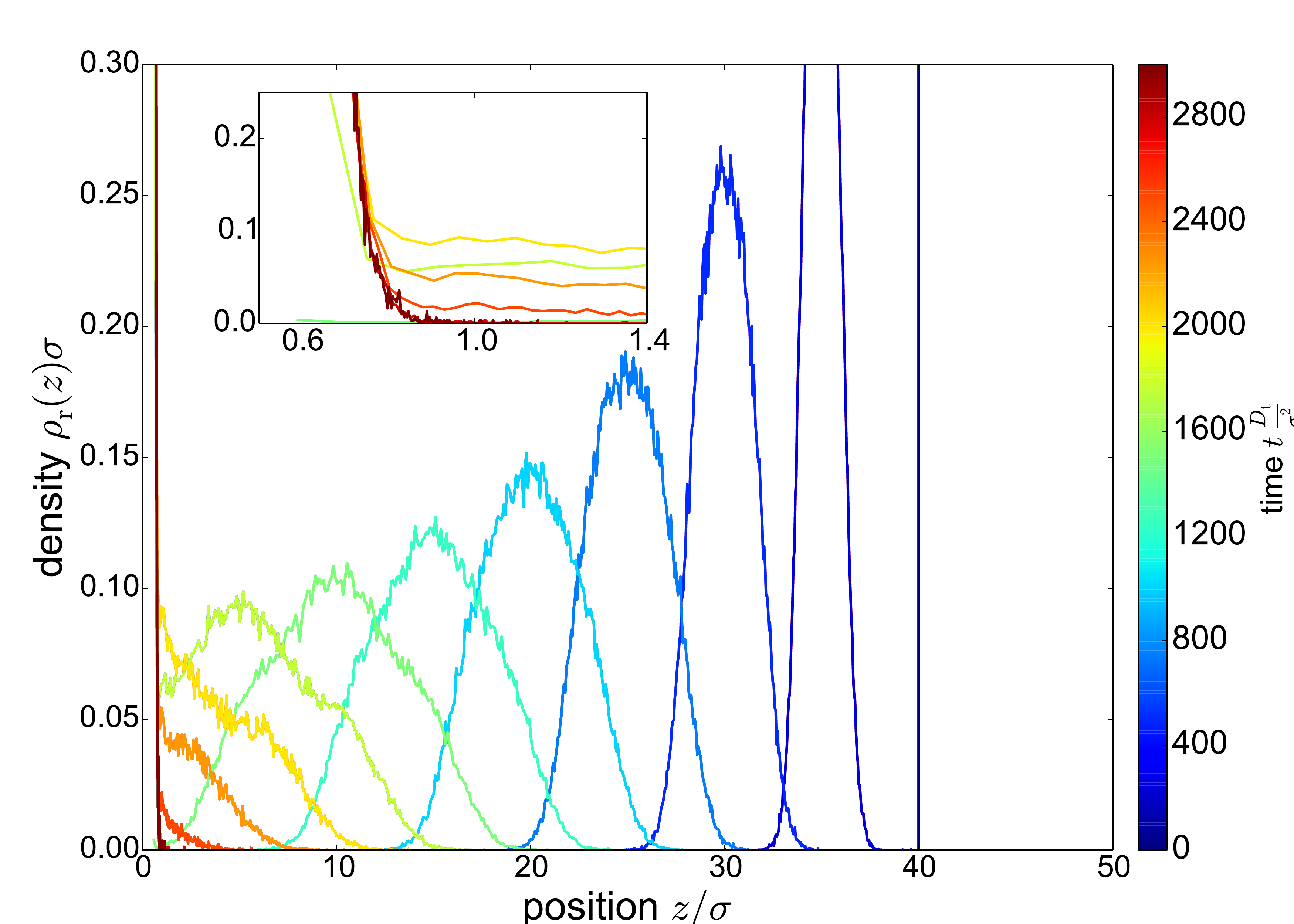}
\caption{Sedimentation profile from simulations. Dependence of the density $\rho_{\mathrm{r}}(z)$ on the position $z$ at different times from simulations of $N=1000$ active particles in a cubic box of linear size $L=50\sigma$ with a wall on bottom ($z=0$) and with gravity pointing in negative $z$-direction. Results are averaged over $10^4$ independent simulations.
At $t=0$, the active particles are randomly placed on a plane located at $z/\sigma=40$.  Different curves correspond to different moments during the time evolution. The model parameters are $v_{s}/v_{g} = 0.2$.}\label{Manysedimen}
\end{figure}

We integrate eqs.~\eqref{eq:simulation-1}-\eqref{eq:simulation-2} with a discretization scheme based on the Euler--Maruyama algorithm in which we take into account the issue of  multiplicative noise. 
We solve eq.~\eqref{eq:simulation-1}-\eqref{eq:simulation-2} in a domain of volume $V=L^3$, with $L=50\sigma$, with a reflective wall at the bottom at $z=0$, and gravity pointing in negative $z$-direction. 
The filling fraction of our system is $\phi=N\frac{\pi}{6}\sigma^3/V=10^{-3}$.
Our results shown below are averaged over $10^4$ independent simulations.
 Figure~\ref{Manysedimen} shows the dependence of the density profile $\rho_{\mathrm{r}}(z,t)$ on the position $z$ at different times. At $t=0$, the active particles are randomly placed on a plane located at $z/\sigma=40$. After some time, all the active particles sediment on the bottom wall. We observe a qualitative agreement of our simulation with our theory. We conclude that, as long as hydrodynamic interactions are weak, or the system is diluted enough, the theory derived for a single active particle is also applicable  to a many-particle system.

\begin{figure}[h!]
\centering
\includegraphics[width=1.0\columnwidth]{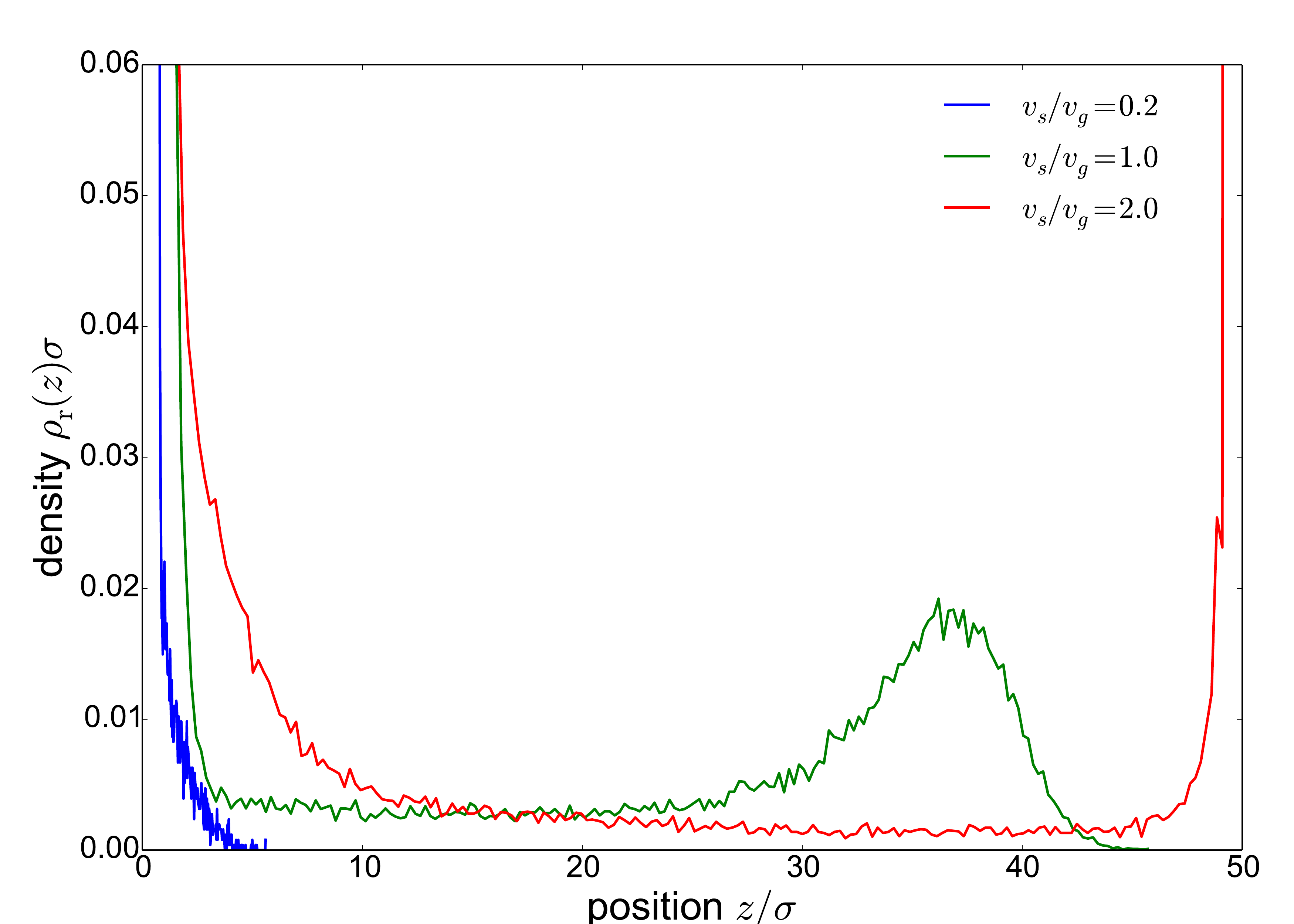}
\caption{Variation of the activity of the particles. Dependence of the density $\rho_{\mathrm{r}}(z)$ on the position $z$ for a system with $N=1000$ active particles, for large $t$. Different curves correspond to different values of $v_{s}/v_{g}$.}\label{tinfpeclet}
\end{figure}

To measure the importance of the active motion with respect to the diffusion, we vary the P\'eclet number $\mathcal{P}$ by changing the self-propulsion $v_{s}$.  Figure \ref{tinfpeclet} shows the results of our simulations for the density profile at large $t$, and   at  different values of self-propulsion $v_{s}$.
 When the activity is lower than the sedimentation velocity, $v_{s}/v_{g}=0.2$, we observe a clear sedimentation profile. If  $v_{s}/v_{g}=1.0$, we observe a weaker sedimentation profile and a peak appears close to the upper part of the simulated domain. This peak is due to balance of the weak, effective hydrodynamic interactions introduced in our simulations with the self-propulsion and the effect of gravity. 
 This is a consequence of the  emerging polar order in sedimenting active particles as discussed in sec.~\ref{sec:analytic}.
 
  Finally, as soon as  $v_{s}>v_{g}$,  we observe an accumulation of particles both on the bottom and top end, which differs from the classical sedimentation profile. We clearly highlight the importance of the activity of the particles, which allows them to move against an external force, in our case the gravity. From a biological point of view, this capacity play an important role, {\it e.g.} the algae need light to survive and they need to move against gravity to reach the surface.

\section{Conclusion}\label{sec:concl}
The dynamics of sedimenting active particles prove to be an interesting arena where different nonequilibrium effects are at play, providing a testbed for our understanding of far-from-equilibrium phenomena.  Including a self-propulsion to the motion of sedimenting colloidal particles ushers in a wealth of intriguing effects unimaginable from the classical results of Perrin~\cite{Perrin-AnnChimPhys-1909}. This is in fact reflected in the considerable interested elicited by this problem~\cite{Palacci-PRL-2010,Ginot-PRX-2015,Tailleur-EPL-2009,Enculescu-PRL-2011,Wolff-EPJE-2013,hermann-softmatter-2018}.

We study the sedimentation process of active Brownian particles in three dimensions. Firstly, we develop an analytical method describing the sedimentation profile of one active particle. We solve analytically the Fokker--Planck equation for an active particle in the presence of gravity and a confining wall at the bottom.  
We address the time evolution of the monopole, and find a solution which matches the late-time density profile in~\cite{Palacci-PRL-2010,Ginot-PRX-2015}.  Furthermore, we calculate the 
the dipolar term, and find the emergence of polar order at the bottom wall, with an accumulation of particles moving against the wall, and a depletion of particles moving away from it. 

Imposing the no-flux condition at the confining bottom wall produces the steady-state solution. This solution is consistent with a number of previous results (most recently~\cite{hermann-softmatter-2018}, for example).

We recover the following experimental results found in~\cite{Palacci-PRL-2010,Ginot-PRX-2015}: (i) the exponentially decaying density profile for the long-time regime and the steady state (ii) the increasing sedimentation length upon increase of the effective diffusivity. Importantly, our method retains the temporal dynamics of the sedimentation process, and therefore in addition to the steady state we also have access to the intermediate states. Our method also allows us to keep the coupling between the rotational and the positional degrees of freedom. In order to characterize more realistic conditions for the sedimentation process, we also consider many particles with weak, effective hydrodynamic interactions and we carry out Brownian dynamics simulations. We are able to measure the importance of the active motion at large times by varying the P\'eclet number $\mathcal{P}$. We recover the density profile, shown in fig.~\ref{tinfpeclet}, found experimentally in ~\cite{Ginot-PRX-2015} and in numerical simulations ~\cite{Wolff-EPJE-2013} of active bottom-heavy particles. However, our model allows us to characterize more in details the richness of the sedimentation process of active particles as function of the activity. Furthermore, the sedimentation profile predicted by our analytical method for one active particle (see fig.~\ref{diffcoefstrength1}) matches our simulations for many particles (see fig.~\ref{Manysedimen}). 

\appendix

\section{Furutsu--Novikov--Donsker relation} \label{sec:app1}

To derive the Fokker--Planck equation, we consider the derivative of  $P(\bm{r},\bm{e},t) = \langle\delta(\bm{r}(t)-\bm{r})\delta(\bm{e}(t)-\bm{e})\rangle$ with respect to time
\begin{align}
\frac{\partial}{\partial t}P(\bm{r},\bm{e},t) &= -\left( v_{s}\bm{e} - v_g\bm{z}\right)\cdot \nabla P(\bm{r},\bm{e},t) \nonumber\\
&-\nabla\cdot\langle\bm{\xi}(t)\delta(\bm{r}(t)-\bm{r})\delta(\bm{e}(t)-\bm{e})\rangle \nonumber\\
			    &- \nabla_{\bm{e}}\cdot\langle[\bm{\xi}_{\mathrm{e}}(t)\times \bm{e}]
\delta(\bm{r}(t)-\bm{r})\delta(\bm{e}(t)-\bm{e})\rangle\,,
\end{align}
where $\nabla_{\bm{e}}\equiv (\frac{\partial}{\partial e_x},\frac{\partial}{\partial e_y}, \frac{\partial}{\partial e_z})^\mathsf{T}$, where the superscript $\mathsf{T}$ indicates transposition. To calculate the ensemble averages involving the noise $\xi$ and $\xi_{e}$, we use the Furutsu--Novikov--Donsker relation~\cite{Klyatskin,Konotop,Frank-PRE-2005,Novikov-Paper-1964, zinnjustin2002,Sevilla-PRE-2014,Sevilla-PRE-2015,Sevilla-PRE-2016}
\begin{equation}
\langle\xi(t)R[\xi]\rangle = \displaystyle\int\limits_{-\infty}^{+\infty}dt'\langle\xi(t)\xi(t')\rangle\langle\frac{\delta R[\xi]}{\delta\xi(t')}\rangle\,,
\label{eq:Novikov}
\end{equation}
where $R[\xi]$ is an arbitrary functional of $\xi$. Physically, relation \eqref{eq:Novikov} helps obtain the dependence of a stochastic observable (e.g. the position of a colloidal particle) on the properties of the noise term.\\

\section{Eigenfunction expansion} \label{sec:app2}

Inserting eq.~\eqref{eq:gen-expansion} into eq.~\eqref{eq:FPEFourier}, and then multiplying by the complex conjugate of the spherical harmonics $Y_{n^\prime}^{*m^\prime }$ and integrating over the solid angle $d\Omega=\sin\theta d\theta d\phi$, we find 
\begin{align}\label{eq:hierarchy-coeff}
\frac{\partial}{\partial t}\widehat{P}_n^m(\bm{k},t)&=- \sum_{n^\prime=0}^\infty \sum_{m^\prime=-n^\prime}^{+n^\prime}\widehat{P}_{n^\prime}^{m^\prime}(\bm{k},t) e^{-D_e (\lambda_{n^\prime} -\lambda_n)t}\nonumber\\
&\int d\Omega \;Y_{n^\prime}^{m^\prime }(\bm{e})\; (iv_s\bm{e}\cdot\bm{k})\;Y_n^{*m}(\bm{e})\,.
\end{align}

To proceed we define the following integrals
\begin{align}
\mathcal{J}_x{}^{m,m'}_{n,n'} & =\int d\Omega\; Y_{n^\prime}^{m^\prime } (\theta,\phi) \sin\theta\cos\phi\; Y_n^{*m}(\theta,\phi)\,,\\
\mathcal{J}_y{}^{m,m'}_{n,n'}& =\int d\Omega\; Y_{n^\prime}^{m^\prime } (\theta,\phi) \sin\theta\sin\phi\; Y_n^{*m}(\theta,\phi)\,,\\
\mathcal{J}_z{}^{m,m'}_{n,n'} & =\int d\Omega\; Y_{n^\prime}^{m^\prime } (\theta,\phi) \cos\theta\; Y_n^{*m}(\theta,\phi)\,.
\end{align}
 Equation~\eqref{eq:hierarchy-coeff} can then be written as
\begin{align}
\frac{\partial}{\partial t}\widehat{P}_n^m(\bm{k},t)&=-iv_s\sum_{n^\prime=0}^\infty \sum_{m^\prime=-n^\prime}^{+n^\prime}\widehat{P}_{n^\prime}^{m^\prime}e^{-D_e (\lambda_{n^\prime} -\lambda_n)t}\nonumber\\
&\left[k_x\mathcal{J}_x{}^{m,m'}_{n,n'}+  k_y\mathcal{J}_y{}^{m,m'}_{n,n'}+  k_z\mathcal{J}_z{}^{m,m'}_{n,n'}\right]\,.
\end{align}

The calculation of the integrals $\mathcal{J}_i{}^{m,m'}_{n,n'}$ is straightforward \cite{Sevilla-PRE-2016,coffey-langevin-2012}. Equation~\eqref{eq:hierarchy-coeff} becomes
\begin{widetext}
\begin{align}
\frac{\partial}{\partial t}\widehat{P}_n^m &= \frac{v_s}{2}e^{-2D_e (n+1)t}
\Bigg\{ (k_y-ik_x) \widehat{P}_{n+1}^{m+1} \left[\frac{(n+m+2)(n+m+1)}{(2n+3)(2n+1)}  \right]^{\tfrac{1}{2}}
- 2ik_z \widehat{P}_{n+1}^{m}  \left[\frac{(n+m+1)(n-m+1)}{(2n+3)(2n+1)}  \right]^{\tfrac{1}{2}}  \nonumber\\
& + \widehat{P}_{n+1}^{m-1}  \left[\frac{(n-m+2)(n-m+1)}{(2n+3)(2n+1)}  \right]^{\tfrac{1}{2}}(k_y+ik_x)\Bigg\}
-\frac{v_s}{2}e^{2D_e nt} \Bigg\{ (k_y-ik_x)\widehat{P}_{n-1}^{m+1}  \left[\frac{(n-m)(n-m-1)}{(2n+1)(2n-1)}  \right]^{\tfrac{1}{2}}  \nonumber\\       &  + 2ik_z \widehat{P}_{n-1}^{m}  \left[\frac{(n+m)(n-m)}{(2n+1)(2n-1)}  \right]^{\tfrac{1}{2}}  + (k_y+ik_x)\widehat{P}_{n-1}^{m-1}  \left[\frac{(n+m)(n+m-1)}{(2n+1)(2n-1)}  \right]^{\tfrac{1}{2}}      \Bigg\}\,.
\end{align}
\end{widetext}

We are interested in the dynamics along the direction of gravity, the $z$-direction; hence we specialize the previous equation to this case. The equation for the coefficients $\widehat{P}_n^m$ specialized to the $z$-direction reads
\begin{align}
\frac{\partial}{\partial t}\widehat{P}^{m}_{n} &= -v_{s}ik_z \Bigg\{ e^{-2D_{\mathrm{e}}(n+1)t} \widehat{P}^{m}_{n+1}\times     \nonumber\\
&\left[\frac{(n+m+1)(n-m+1)}{(2n+3)(2n+1)} \right]^{\tfrac{1}{2}}+\nonumber\\
&e^{2D_{\mathrm{e}}nt}\widehat{P}^{m}_{n-1}\left[\frac{(n+m)(n-m)}{(2n+1)(2n-1)} \right]^{\tfrac{1}{2}} \Bigg\}\,.
	\label{Pz}	      
\end{align}

\section{Telegrapher's equation}\label{sec:app3}

Here we provide details of the computation of the telegrapher's eq. \eqref{eq:Telegrapher}. We start by considering the equations for the two coefficients $\widehat{P}^{0}_{0}$ and $\widehat{P}^{0}_{1}$
\begin{align}
\frac{\partial}{\partial t}{\widehat{P}}^{0}_{0} &= -\frac{v_{s}}{\sqrt{3}}e^{-2D_{\mathrm{e}}t}ik_{z}{\widehat{P}}^{0}_{1}\,,\\
\frac{\partial}{\partial t}{\widehat{P}}^{0}_{1} &= -{v_{s}}e^{-4D_{\mathrm{e}}t}\sqrt{\frac{4}{15}}ik_{z}\widehat{P}^{0}_{2} -\frac{v_{s}}{\sqrt{3}}e^{2D_{\mathrm{e}}t}ik_{z}{\widehat{P}}^{0}_{0}\,,\label{eq:der-p10}
\end{align}
combining these two Eqs. yields
\begin{equation}
\frac{\partial^2}{\partial t^2}{\widehat{P}}^{0}_{0} = 2D_{\mathrm{e}}\frac{v_{s}}{\sqrt{3}}e^{-2D_{\mathrm{e}}t}ik_{z}{\widehat{P}}^{0}_{1}-\frac{v_{s}}{\sqrt{3}}e^{-2D_{\mathrm{e}}t}ik_{z}\frac{\partial}{\partial t}{\widehat{P}}^{0}_{1}\,,
\end{equation}
and after replacing the last term on the right hand side with eq.~\eqref{eq:der-p10} we find
\begin{equation*}
\frac{\partial^2}{\partial t^2}{\widehat{P}}^{0}_{0} + 2D_{\mathrm{e}}\frac{\partial}{\partial t}{\widehat{P}}^{0}_{0} + \frac{v_{s}^{2}}{3}k_{z}^{2}{\widehat{P}}^{0}_{0} = - v_{s}^{2}\sqrt{\frac{4}{45}}k_{z}^{2}{\widehat{P}}^{0}_{2} e^{-6D_{\mathrm{e}}t}\,.
\end{equation*}
Finally, neglecting the higher order yields
\begin{equation*}
\frac{\partial^2}{\partial t^2}{\widehat{P}}^{0}_{0} + 2D_{\mathrm{e}}\frac{\partial}{\partial t}{\widehat{P}}^{0}_{0} + \frac{v_{s}^{2}}{3}k_{z}^{2}{\widehat{P}}^{0}_{0} = 0\,,
\end{equation*}
which is the telegrapher's eq.~\eqref{eq:Telegrapher}.\\

\section{{\color{black}Monopole reflective boundary}}\label{sec:app4}
{\color{black} Let's start from the solution of a diffusion process 
\[
\rho(z,t) = \frac{e^{-(z-a-v_gt)^2/4Dt}}{\sqrt{4\pi D_tt}}\,.
\]
It is convenient to introduce a change in the independent variable \cite{chandrasekhar-1943-stochastic}
\[
\rho(z,t)=Ue^{\frac{v_g(z-a)}{2D_t}-\frac{v_g^2t}{4D_t}}\,,
\]
where
\[
U=\frac{1}{\sqrt{4\pi D_tt}}e^{-(z-a)^2/4D_tt}\,.
\]
Finally
\begin{align*}
\rho(z,t)=Ue^{\frac{v_g(z-a)}{2D_t}-\frac{v_g^2t}{4D_t}}=\frac{1}{\sqrt{4\pi D_tt}}e^{-(z-a-v_g t)^2/4D_tt}
\end{align*}
In order to take into of the reflecting barrier, $\rho(r,t)$ becomes \cite{sommerfeld-1949-partial,aleksandrov-1999-mathematics,cox-2017-theory,chandrasekhar-1943-stochastic}
\begin{equation}
\rho_\mathrm{r}(z,t) = \rho(z,t|a)+A\rho(z,t|-a)+\int\limits^{-a}_{-\infty}k(\xi)\rho(z,t|\xi)d\xi\,,
\label{densitywall}
\end{equation}
which tells us that an isolated point (image) $z=-a$ is not sufficient, but we need a continuous sequence of images which take place at all points $\xi<-a$.
The Robin boundary condition for our system reads
\begin{align*}
&\left[D_t\frac{\partial}{\partial z}\rho_\mathrm{r}(z,t)-v_g\rho_\mathrm{r}(z,t)\right]_{z=0}=0\\
&\iff \left[D_t\frac{\partial}{\partial z}U_\mathrm{r}-\frac{1}{2}v_gU_\mathrm{r})\right]_{z=0}=0
\,.\label{app_boundary}
\end{align*}
then
\begin{align*}
\sqrt{4\pi D_tt}U_r &= e^{-(z-a)^2/4D_tt} + Ae^{-(z+a)^2/4D_tt} \\
&+\int\limits_{-\infty}^{-a}k(\xi)e^{-(z-\xi)^2/4D_tt}d\xi\,.
\end{align*}
By applying the Robin boundary condition \eqref{densitywall} reads
\begin{align}
&\frac{a}{2t}e^{-a^2/4D_tt}(1-A)-D_tk(-a)e^{-a^2/4D_tt}\nonumber\\
&+D_t\int\limits_{\infty}^{-a}\frac{\partial}{\partial\xi}k(\xi)e^{-\xi^2/4D_tt}d\xi-\frac{1}{2}v_ge^{-a^2/4D_tt}(1+A)\nonumber\\
&-\frac{1}{2}v_g\int\limits_{\infty}^{-a}k(\xi)e^{-\xi^2/4D_tt}d\xi = 0
\end{align}
By setting the terms of different time dependance individually equal to zero, the coefficients $A$ and $k(\xi)$ read
\begin{itemize}
\item $A=1$ 
\item $k(-a)=-\frac{v_g}{D_t}$
\item $k(\xi)=-\frac{v_g}{D_t}e^{\frac{2v_g(\xi +a)}{D_t}}$
\end{itemize}
By replacing the function and coefficient inside the equation, the solution for $U_r$ is given by
\begin{align*}
\sqrt{4\pi D_tt}U_r &= e^{-(z-a)^2/4D_tt} + e^{-(z+a)^2/4D_tt} \\
&-\int\limits_{-\infty}^{-a}\frac{v_g}{D_t}e^{\frac{2v_g(\xi +a)}{D_t}}e^{-(z-\xi)^2/4D_tt}d\xi\,,
\end{align*}
and by rewriting the integral term
\begin{align*}
&-\int\limits_{-\infty}^{-a}\frac{2v_g}{D_t}e^{\frac{v_g(\xi +a)}{D_t}}e^{-(z-a)^2/4D_tt}d\xi \\
&= \frac{2v_g}{D_t}\int\limits^{\infty}_{a}e^{\frac{v_g(\xi -a)}{D_t}}e^{-(z+\xi)^2/4D_tt}d\xi\\
&=\frac{v_g}{D_t}\int\limits^{+\infty}_{a}e^{\frac{v_g(\xi -a)}{2D_t}}e^{-(\xi-a)^2/4D_tt}d\xi
\end{align*}
and by doing a change of variable \cite{chandrasekhar-1943-stochastic} for the limits of the integral $U_r$ reads
\begin{align*}
U_r&= \frac{1}{\sqrt{4\pi D_tt}}\left[e^{-(z-a)^2/4D_tt}+e^{-(z+a)^2/4D_tt}\right]\\
&+\frac{v_g}{D_t\sqrt{\pi}}e^{(\frac{v_g^2t}{4D_t}-\frac{v_g(z+a)}{2D_t})}\int\limits_{\frac{z+a-v_gt}{2\sqrt{D_tt}}}^{+\infty}e^{-\eta^2}d\eta\,.
\end{align*}
Finally, the complete solution reads
\begin{align}
\rho_\mathrm{r}(z,t) &= U_re^{\frac{v_g(z-a)}{2D_t}-\frac{v_g^2t}{4D_t}}\nonumber\\
&=\frac{1}{\sqrt{4\pi D_tt}}\left[e^{-(z-a)^2/4Dt}+e^{-(z+a)^2/4D_tt}\right]e^{(\frac{v_g(z-a)}{2D_t}-\frac{v_g^2t}{4D_t})}\nonumber\\
&+\frac{v_g}{D_t\sqrt{\pi}}e^{-v_gz/D_t}\int\limits_{\frac{z+a-v_gt}{2\sqrt{D_tt}}}^{+\infty}e^{-\eta^2}d\eta\,.
\end{align}
}

\section{Probability density function}\label{sec:app5}
We want to take into account the polarization $D(z,t)$, in the probability density function $P(z,\cos(\theta),t)$, defined in Fourier space as 
\[
D(k_z,t)=\sqrt{\frac{3}{4\pi}}e^{-(iv_gk_z-D_tk_z^2)t}e^{-iak_z}\widehat{P} ^0_1\,.
\]
We can find $\widehat{P}^0_1$ by using eq.(\ref{Pz}). From there, we can work out the associated telegrapher's equation and by neglecting the higher orders, we find
\[
\frac{\partial^2}{\partial_t^2}\widehat{P}^0_1-(4\sqrt{\frac{4}{15}}-\frac{2}{3})
v_s^2k_z^2D_e\widehat{P}^0_1 = 0\,.
\]
Simple computations yield
\[
\widehat{P}^0_1 =C_1e^{fv_sk_zD_et}+C_2e^{-fv_sk_zD_et}\,,
\]
where $f=\sqrt{4\sqrt{\frac{4}{15}}-\frac{2}{3}}$. By applying the inverse Fourier transform
\begin{align*}
D(z,t)&=\sqrt{\frac{3}{8\pi^2}}\left[\frac{C_1}{\sqrt{2tD_e}}e^{-(a-z+v_gt+fitD_ev_s)^2/4tD_t}\right.\\
&+\left.\frac{C_2}{\sqrt{2tD_e}}e^{-(a-z+v_gt-fitD_ev_s)^2/4tD_t}\right]\,.
\end{align*}

We only consider the two first terms of the probability density function
\[
P(z,\cos(\theta),t) \simeq \rho(z,t) + D(z,t)\cos(\theta)
\]
and by plugging in the polarization, and focusing only on the real part of the exponentials, the probability density function reads 
\begin{align}
&P(z,\cos(\theta),t) = \frac{1}{\sqrt{8\pi}} \Bigg[e^{-2D_{\mathrm{e}}t} \frac{\widetilde{F}}{\sqrt{D_\mathrm{eff}^- t}}e^{-{(z-a-v_{g}t)^{2}}/({4 D_\mathrm{eff}^- t})}\nonumber\\
&+\frac{\widetilde{G}}{\sqrt{D_\mathrm{eff}^+ t}}e^{-{(z-a-v_{g}t)^{2}}/({4D_\mathrm{eff}^+ t})}\Bigg]\nonumber\\
&+\sqrt{\frac{3}{8\pi^2}}\frac{\cos(\theta)}{\sqrt{2tD_t}}
e^{-(a^2-2az+z^2+2atv_g-2tzv_g+t^2v_g^2-f^2t^2D_e^2v_s^2)/2D_t}\nonumber\\
&\times\big[C_1+C_2\big]\cos(\alpha)\,,
\end{align}
where 
\begin{align*}
\alpha &= \frac{-afD_ev_s+fzD_ev_s-ftD_ev_gv_s}{2D_t}\,.
\end{align*}
In order to find the coefficients $C_1$ and $C_2$,  we apply the Robin boundary conditions at $z=0$, $J_z = 0$, to eq.(\ref{eq:continuity}). Moreover, $\widetilde{F}=e^{2D_et}$ and $\widetilde{G}=1$ to keep the mass constant over time. After some computations we find
\begin{equation}
C_1 = \frac{C_2B_1+B_2}{C}\,,
\end{equation}
where
\begin{align}
B_1 &= -(-v_s+v_g)\frac{-1}{4\sqrt{tD_t}}\sqrt{\frac{3}{\pi}}e^{\Lambda}\cos(\Omega)\nonumber\\
&-D_t\sqrt{\frac{3}{\pi}}\frac{\cos(\pi)}{8tD_t\sqrt{tD_t}}e^{\Lambda}\cos(\Omega)\nonumber\\
&-\frac{-1}{4\sqrt{tD_t}}\sqrt{\frac{3}{\pi}}e^{\Lambda}fD_ev_s\sin(\Omega)\,,\\
\Omega &= \frac{faD_ev_s+ftD_ev_gv_s}{2D_t}\,,\\
\Lambda &= \frac{-a^2-2atv_g-t^2v_g^2+f^2t^2D_e^2v_s^2}{4tD_t}\,,
\end{align}
and
\begin{align}
B_2 &= -(-v_s+v_g)\frac{1}{\sqrt{8\pi}} \Bigg[\frac{1}{\sqrt{D_\mathrm{eff}^- t}}e^{-{(z-a-v_{g}t)^{2}}/({4 D_\mathrm{eff}^- t})}\nonumber\\
&+\frac{1}{\sqrt{D_\mathrm{eff}^+ t}}e^{-{(z-a-v_{g}t)^{2}}/({4D_\mathrm{eff}^+ t})}\Bigg]\nonumber\\
&-D_t\frac{(a+v_gt)}{2\sqrt{2\pi}}\bigg[\frac{e^{(a+v_gt)^2/(4 D_\mathrm{eff}^- t)}}{2tD_\mathrm{eff}^- \sqrt{t D_\mathrm{eff}^-}}+\frac{e^{(a+v_gt)^2/(4 D_\mathrm{eff}^+ t)}}{2tD_\mathrm{eff}^+ \sqrt{t D_\mathrm{eff}^+}}\bigg]\,,
\end{align}
and
\vspace{-.5cm}
\begin{align}
C &= (-v_s+v_g)\frac{-1}{4\sqrt{tD_t}}\sqrt{\frac{3}{\pi}}e^{\Lambda}\cos(\varphi)\nonumber\\
&+D_t\bigg[-\sqrt{\frac{3}{\pi}}\frac{(a+v_gt)}{8tD_t\sqrt{tD_t}}e^{\Lambda}\cos(\varphi)\nonumber\\
&-\sqrt{\frac{3}{\pi}}\frac{-1}{4t\sqrt{tD_t}}e^{\Lambda}fD_ev_s\sin(\varphi)\bigg]\,,\\
\varphi &= \frac{-faD_ev_s-ftD_ev_gv_s}{2D_t} \,.
\end{align}
Finally, the probability density function is given by
\begin{align}
&P(z,\cos(\theta),t) = \frac{1}{\sqrt{8\pi}}\Bigg[\frac{1}{\sqrt{D_\mathrm{eff}^- t}}e^{-{(z-a-v_{g}t)^{2}}/({4 D_\mathrm{eff}^- t})}\nonumber\\
&+\frac{1}{\sqrt{D_\mathrm{eff}^+ t}}e^{-{(z-a-v_{g}t)^{2}}/({4D_\mathrm{eff}^+ t})}\Bigg]\nonumber\\
&+\sqrt{\frac{3}{4\pi}}\frac{\cos(\theta)}{\sqrt{2tD_t}}\Bigg[C_2\cos(\alpha)\Big(\frac{B_1}{C}+1\Big)+\frac{B_2}{C}\cos(\alpha)\Bigg]
\nonumber\\
&\times e^{-(a^2-2az+z^2+2atv_g-2tzv_g+t^2v_g^2
-ft^2D_e^2v_s^2)/(4D_t t)}.
\end{align}
For the sake of simplicity we set $C_2=1$.

\section*{Acknowledgements}

We gratefully acknowledge Soumyajyoti Biswas,  Rebekka Breier, Stephan Herminghaus, Hao Chen and Michael Wilczek for helpful conversations. We thank the Max Planck Society and the Max Planck Center for Complex Fluid Dynamics for funding. M.G.M. gratefully acknowledges support from the Deutsche Forschungsgemeinschaft (SFB 937, project A20).\\

\section*{Author contributions statement}

M.G.M. conceived the project. J.V implemented the theory and performed simulations. M.G.M. and J.V interpreted the data and wrote the paper. All  authors discussed the results and commented on the manuscript.



%



\end{document}